\documentclass[a4paper,aps,pra,showpacs,superscriptaddress,nofootinbib,twocolumn]{revtex4-1}      
\usepackage[utf8]{inputenc}  
\usepackage[T1]{fontenc}     
\usepackage[british]{babel}  
\usepackage{lmodern}  
\usepackage[scaled=0.86]{berasans}  
\usepackage[scaled=1.03]{inconsolata} 
\usepackage[usenames,dvipsnames]{color} 
\usepackage[colorlinks,citecolor=blue,linkcolor=magenta,urlcolor=blue]{hyperref}  
\usepackage{graphicx} 
\usepackage[babel]{microtype}  
\usepackage{amsmath,amssymb,amsthm,bm,mathtools,amsfonts,mathrsfs,bbm} 
\usepackage{xspace}  
\usepackage{pgfplots}  
\DeclareMathOperator{\tr}{tr}
\DeclareMathOperator{\Tr}{Tr}

\newcommand{\bra}[1]{\left\langle #1 \right|}
\newcommand{\ket}[1]{\left| #1 \right\rangle}

\newcommand{\ketbra}[2]{\left|#1\middle\rangle\middle\langle#2\right|}

\newcommand{\eg}{\emph{e.g.}\@\xspace}
\newcommand{\ie}{\emph{i.e.}\@\xspace}

\newcommand{\ba}{\begin{eqnarray}}
\newcommand{\ea}{\end{eqnarray}}
\newcommand{\ban}{\begin{eqnarray*}}
\newcommand{\ean}{\end{eqnarray*}}
\newcommand{\ave}[1]{\langle#1\rangle}

\begin{document}

\title{Local hidden variable models for entangled quantum states using finite shared randomness}
\hypersetup{pdftitle={{Simulating the correlations of entangled quantum states with finite resources}}}

\author{Joseph Bowles}\thanks{These authors contributed equally to this work.}
\affiliation{D\'epartement de Physique Th\'eorique, Universit\'e de Gen\`eve, 1211 Gen\`eve, Switzerland}

\author{Flavien Hirsch}\thanks{These authors contributed equally to this work.}
\affiliation{D\'epartement de Physique Th\'eorique, Universit\'e de Gen\`eve, 1211 Gen\`eve, Switzerland}

\author{Marco T\'ulio Quintino}
\affiliation{D\'epartement de Physique Th\'eorique, Universit\'e de Gen\`eve, 1211 Gen\`eve, Switzerland}

\author{Nicolas Brunner}
\affiliation{D\'epartement de Physique Th\'eorique, Universit\'e de Gen\`eve, 1211 Gen\`eve, Switzerland}

\date{\today}  

\begin{abstract}
The statistics of local measurements performed on certain entangled states can be reproduced using a local hidden variable (LHV) model. While all known models make use of an infinite amount of shared randomness---the physical relevance of which is questionable---we show that essentially all entangled states admitting a LHV model can be simulated with finite shared randomness. Our most economical model simulates noisy two-qubit Werner states using only $\log_2(12)\simeq 3.58$ bits of shared randomness. We also discuss the case of POVMs, and the simulation of nonlocal states with finite shared randomness and finite communication. Our work represents a first step towards quantifying the cost of LHV models for entangled quantum states.
\end{abstract}

\maketitle

\section{Introduction}

Quantum systems exhibit a wide range of non-classical and counter-intuitive phenomena, such as quantum entanglement \cite{horodecki_review} and Bell nonlocality \cite{bell64,brunner_review}. In recent years, a great effort has been devoted to understanding the relation between entanglement and nonlocality; see \cite{brunner_review}. While entanglement is necessary to demonstrate nonlocality (\ie violation of a Bell inequality), it is not yet clear whether all entangled states can lead to nonlocality when considering the most general scenario \cite{masanes08,vertesi14}. Nevertheless, entanglement and nonlocality are proven to be different in the simplest scenario in which local (non-sequential) measurements are performed on a single copy of an entangled state. As discovered by Werner \cite{werner89}, there exist entangled states that can provably not violate any Bell inequality, since the state admits a local hidden variable (LHV) model. While Werner focused on projective measurements, Barrett \cite{barrett02} showed that the result holds for the most general non-sequential measurements, so-called positive operator valued measures (POVMs). 

Following these early results, plenty of works have investigated these ideas; see \cite{augusiak_review} for a recent review. LHV models were reported for entangled states with less symmetry than Werner states \cite{almeida07,acin06,hirsch13,bowles14,tran14}. Multipartite states were discussed as well \cite{toth05,augusiak14}. Interestingly, it was shown that in certain cases, the nonlocality of local entangled states can be activated, \eg by considering sequential measurements \cite{popescu95,hirsch13}. More recently, interest was devoted to a special class of LHV models, referred to as local hidden state (LHS) models, which naturally arise in the context of Einstein-Podolsky-Rosen (EPR) steering \cite{wiseman07,reid08}, and essentially require that the local variable represents a quantum state; see \cite{wiseman07} for details, and  \cite{werner89,barrett02,almeida07,bowles14,sania14} for examples of LHS models.

Here we discuss novel types of questions in this context, namely that of \emph{quantifying LHV models}. Specifically, given a local entangled state, we ask what resources are required to construct a LHV model, \ie what is the cost of classically simulating the correlations of the state. As a figure of merit, we consider the minimal dimension of the shared local (hidden) variable that is needed; that is, how much classical information (how many bits) is necessary to encode the local variable. Note that all LHV models constructed so far are maximally costly according to our measure as they make use of shared variables of infinite dimension. Hence such models would require a communication channel of infinite capacity, the physical relevance of which is questionable. For instance, in Werner's model, the local variables are unit vectors $\vec{\lambda}$ (\eg vectors on the Bloch sphere). Importantly, although these vectors are of a given dimension, the model requires an infinite number of them, as vectors $\vec{\lambda}$ are taken from the uniform distribution over the sphere.    

Hence, a natural question is whether it would be in fact possible to simulate the correlations of an entangled state using shared variables of finite dimension (\ie a finite number of shared random bits). Here we show that essentially any entangled state admitting a LHV model can be simulated with finite shared randomness, considering arbitrary local projective measurements. We discuss in detail the case of Werner states of two-qubits. We also show that the simulation of arbitrary POVMs on certain entangled states is possible using finite shared randomness. Finally, we consider the simulation of nonlocal entangled states (\ie which can violate a Bell inequality), in which case communication between the parties is necessary. In particular, we show that the simulation of any full rank entangled state can be achieved using only finite communication.

Our work provides a perspective on understanding how the correlations of local entangled states differ from those of fully separable states. On the one hand, it shows that there is no fundamental difference between the two cases, in the sense that finite shared randomness is enough for both (at least for certain entangled states). Recall that the correlations of separable states can always be simulated using $2 \log_2(d)$ bits \cite{zyczkowski06}, where $d$ denotes the local Hilbert space dimension of the state. On the other hand, our results suggest that the simulation of entangled states is in general more costly compared to that of separable states---despite the fact that both classes of states can never lead to Bell inequality violation.

\section{Preliminaries}

We consider a bipartite Bell scenario. Two distant observers, Alice and Bob, share a quantum state $\rho$ (of Hilbert space dimension $d \times d$) and perform local measurements $A = \{ A _a \} $ and $ B = \{ B_b \}$, respectively. The observed statistics are local (in the sense of Bell), if they can be decomposed as follows \cite{bell64,brunner_review}:
\ba 
\label{localprob} \Tr ( A_a \otimes B_b \; \rho ) &= \int \pi(\lambda) \; p_A (a|A,\lambda ) \; p_B (b|B,\lambda) \; d\lambda 
\ea
where $\lambda$ represents a shared (hidden) variable, distributed according to density $\pi(\lambda)$. If a decomposition of the form \eqref{localprob} exists for all possible local measurements, we say that the state $\rho$ is local as it will never violate any Bell inequality. The LHV model is then characterized by the distributions $ \pi(\lambda)$, and $ p_A (a|A,\lambda )$, $p_B(b|B,\lambda)$ which are Alice's and Bob's local response functions.

Trivially, any state $\rho$ that is separable is local. Indeed, one can write $\rho = \sum_{\lambda=1}^{d^2} p_\lambda \rho_A^\lambda \otimes \rho_B^\lambda $ \cite{zyczkowski06}. Here the local variable $\lambda$ is distributed according to $p_\lambda$, with $\sum_{\lambda} p_\lambda = 1 $, and the local response functions are simply $p_A (a|A,\lambda ) = \Tr ( A_a \rho_A^\lambda)$ for Alice and similarly for Bob. Note that the shared variable takes only $d^2$ different values here, and can thus be encoded in $2 \log_2(d)$ bits. More interestingly, there exist entangled states $\rho$ which are local. The most famous example is the Werner state, which for the case $d=2$ takes the form
\ba \label{werner}
\rho_W(\alpha) = \alpha \ket{ \psi^{-}}\bra{ \psi^{-}}+(1-\alpha)\mathbb{I}/4
\ea
where $\ket{ \psi^{-}}=(\ket{01}-\ket{10})/\sqrt{2}$ is the singlet state and $\mathbb{I}/4$ is the maximally mixed two-qubit state. After showing that the state $\rho_W(\alpha)$ is entangled for $\alpha>1/3$, Werner \cite{werner89} constructed a local model for arbitrary projective measurements for $\alpha \leq 1/2$; later another local model was constructed for $\alpha\lesssim0.66$ \cite{acin06}. Considering the most general non-sequential measurements, \ie POVMs, a local model was presented for $\alpha \leq 5/12$ \cite{barrett02}. 

A common feature of these local models (and to the best of our knowledge, of all known LHV models) is the fact that the shared variable $\lambda$ takes an infinite number of different values; typically, $\lambda$ denotes a (unit) vector, which is taken randomly from a uniform distribution over the sphere. Hence $\lambda$ requires an infinite number of bits to be encoded, in stark contrast with the case of separable states, where $2\log_2(d)$ bits are enough. Therefore, it is rather natural to ask if this represents a fundamental difference between local entangled states and separable ones. Below we will show that this is not the case, by exhibiting LHV models for entangled states requiring only finite resources, \ie where $\lambda$ can be encoded with a finite number of bits.

\section{Simulating Werner states with finite shared randomness}

We present local models using a finite amount of shared randomness, simulating the correlations of Werner states $\rho_W(\alpha)$ for $\alpha<0.5$ for all projective measurements; extensions to $\alpha \lesssim 0.66$ are given in the next section. Alice and Bob receive here Bloch vectors $\vec{a}$ and $\vec{b}$ (representing observables $A=\vec{a} \cdot \vec{\sigma}$ and similarly for Bob) and should provide outcomes $a,b =\pm1$ such that 
\ba \ave{a}=\ave{b}=0 \quad , \quad  \ave{ab} = - \alpha \vec{a} \cdot \vec{b}. \ea

For clarity, we start by presenting a simple model using only $\log_2(12)$ bits of shared randomness, which works for $\alpha \lesssim 0.43$. 
Our model uses the icosahedron, one of the 5 platonic solids in dimension 3. The icosahedron has 12 vertices represented by the normalized vectors $\vec{v}_{\lambda} \in V$, which satisfy the following properties
\begin{align}
\label{oppvec} &\forall \;\vec{v}_{\lambda} \;\exists\;\vec{v}_{j} \text{ s.t. } \vec{v}_{\lambda}=-\vec{v}_{j} \\
\label{sumhs} &\sum_{j \; s.t. \; \vec{v}_{j}\cdot\vec{v}_{\lambda}\geq0}\vec{v}_{j}=\gamma \vec{v}_{\lambda} \hspace{1cm} \forall \lambda 
\end{align}
with $\gamma=1+\sqrt{5}$. Note that the radius of a sphere inscribed inside the icosahedron is given by $\ell=\sqrt{(5+2\sqrt{5})/15}$. In our model the shared variable $\lambda\in\{1,\ldots,12\}$ is distributed uniformly and represents one of the $12$ vertices of the icosahedron. That is, when Alice and Bob receive $\lambda$, they will use vector $\vec{v}_\lambda$. \\

{\bf Protocol 1.} Alice and Bob share $\lambda\in\{1,\ldots,12\}$, uniformly distributed. Upon receiving setting $\vec{a}$, Alice calculates the subnormalized vector $\vec{a}'=\ell \vec{a}$. This ensures that $\vec{a}'$ lies inside the convex hull of $V$ and so Alice can find a convex decomposition $\vec{a}' = \sum_{i}\omega_{i}\vec{v}_{i}$ with $\sum_{i}\omega_{i}=1$ and $\omega_i\geq0$. Then, with probability $\omega_i$, she outputs $a=\pm1$ with probability $(1 \pm  \text{sgn}[\vec{v}_{\lambda}\cdot\vec{v_i}])/2$. Bob, upon receiving $\vec{b}$, outputs $b=\pm1 $ with probability $(1 \mp \;\vec{b}\cdot\vec{v}_{\lambda})/2$.\\

We now show that the protocol reproduces the desired statistics. We start with the correlator: 
\begin{align} \label{corr}
\langle ab \rangle &=-\frac{1}{12}\sum_{\lambda}\sum_{i}\omega_{i}\;\text{sgn}(\vec{v}_i\cdot\vec{v}_{\lambda})\;\vec{v}_{\lambda}\cdot\vec{b} 
\end{align}
Interchanging the sums, we first calculate 
\ba \sum_{\lambda}\;\text{sgn}(\vec{v}_i\cdot\vec{v}_{\lambda})\;\vec{v}_{\lambda}\cdot\vec{b} = 2\gamma \; \vec{v}_{i}\cdot\vec{b}
\ea
which follows from \eqref{oppvec} and \eqref{sumhs}; details in Appendix A. Inserting the last expression in \eqref{corr}, we get
\ba 
\ave{ab} =  -\frac{\gamma}{6}\sum_{i}\omega_{i}  \; \vec{v}_{i}\cdot\vec{b}   
= -\frac{\ell\gamma}{6}\;\vec{a}\cdot\vec{b} \simeq -0.43\;\vec{a}\cdot\vec{b}.
\ea
Finally, we compute Alice's marginal 
\begin{align} \label{mA}
\langle a \rangle &=-\frac{1}{12}\sum_{\lambda}\sum_{i}\omega_{i}\;\text{sgn}(\vec{v}_i\cdot\vec{v}_{\lambda}) = 0
\end{align}
which can be seen from \eqref{oppvec}. Similarly, we get that $\ave{b}=0$. Therefore, the model simulates $\rho_W(\alpha)$ for $\alpha \simeq 0.43$. Extension to smaller values of $\alpha$ is straightforward.

The above protocol can be adapted to any polyhedron satisfying conditions \eqref{oppvec} and \eqref{sumhs}. Natural candidates are the Platonic solids, except for the tetrahedron which does not satisfy \eqref{oppvec}. Among these, the icosahedron turns out to be optimal here; see Appendix A. Hence, in order to simulate Werner states which are more entangled, \ie going beyond  $\alpha \simeq 0.43$, we need another method.

We now present a protocol, which will allow us to relax condition \eqref{sumhs}. Specifically, we consider again a 3-dimensional polyhedron $V$ with $D$ vertices $\vec{v}_{i}$, but only demand that is satisfy $\eqref{oppvec}$ (which can always be achieved at the expense of doubling the number of vertices of a given polyhedron). As before, the shared variable $\lambda \in \{1,...,D\}$ encodes the choice of vertex, and is uniformly distributed. Having abandoned condition \eqref{sumhs}, we have for each vertex $\vec{v}_{\lambda} $: 
\ba  \label{mod5} 
\sum_{j \; s.t. \;  \vec{v}_{j}\cdot\vec{v}_{\lambda}\geq0}\vec{v}_{j}=\gamma_{\lambda} \vec{m}_{\lambda}
\ea
where $\vec{m}_{\lambda}$ is a normalized vector and generally $\vec{m}_{\lambda}\neq\vec{v}_{\lambda}$. Let us define $\gamma_{\text{min}} = \text{min}_{\lambda} ( \gamma_{\lambda} )$. Note that there are now two polyhedra of interest: (i) $V$, that is defined by the vertices $\vec{v}_{\lambda}$ and (ii) $M$, defined by the vertices $\vec{m}_{\lambda}$, which are in one-to-one correspondence with the $\vec{v}_{\lambda}$. Consider the following protocol.

\begin{figure} \begin{center}
\label{graph}
\includegraphics[scale=0.95]{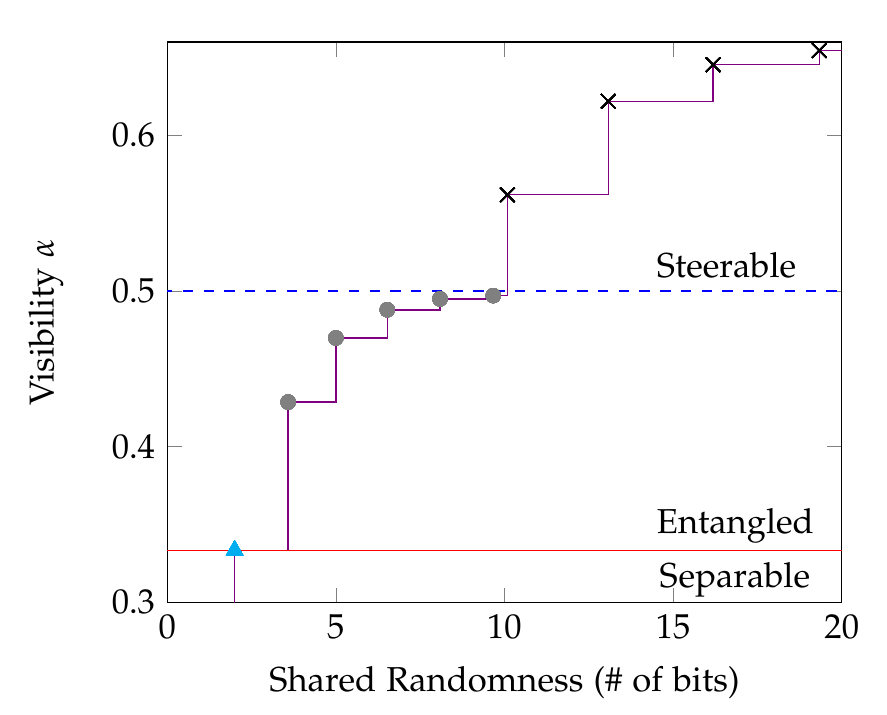}
\caption{Simulation of two-qubit Werner states $\rho_W(\alpha)$ with finite shared randomness. The graph shows the relation between the visibility $\alpha$ (essentially the degree of entanglement) and the amount of shared randomness, quantified in bits. For $\alpha\leq1/3$ (below the solid line) the state is separable, hence 2 bits of shared randomness suffice (triangle). For $1/3 < \alpha \lesssim 0.43$, the state can be simulated with $\log_2(12)$ bits of shared randomness using protocol 1. For $0.43 \lesssim \alpha \lesssim 0.66$, $\rho_W(\alpha)$ can be simulated with a larger (but nevertheless finite) amount of shared randomness. For $0.43 \lesssim \alpha < 0.5$, we have a LHS model (using protocol 2). For $0.5<\alpha\lesssim 0.66$ the state becomes steerable but can nevertheless be simulated by a LHV model using finite shared randomness, by applying Result 1 to the model of Ref. \cite{acin06} (see main text).} 
\end{center}
\end{figure}

{\bf Protocol 2.} Alice and Bob share $\lambda \in \{1,\ldots,D\}$ uniformly distributed. Upon receiving setting $\vec{a}$, Alice calculates the subnormalized vector $\vec{a}'=\ell\vec{a}$ where $\ell$ is the radius of the largest sphere fitting inside $M$ and centred on the origin. This ensures that $\vec{a}'$ lies inside the convex hull of $M$ and Alice can therefore find a convex decomposition $\vec{a}'= \sum_{i=1}^D \omega_{i} \vec{m}_{i}$. Then, with probability $p_i = \omega_i \gamma_{\text{min}} / \gamma_{i} $ she outputs $ a = \text{sgn}(\vec{v}_{i} \cdot \vec{v}_{\lambda}) $, and with probability $(1 - \sum_{i} p_i)$ she outputs a random bit. Bob, upon receiving $\vec{b}$, outputs $b=\pm1 $ with probability $(1 \mp \;\vec{b}\cdot\vec{v}_{\lambda})/2$.

The resulting correlations are given by
\begin{align}
\langle ab \rangle &=-\frac{1}{D} \sum_{\lambda}\sum_{i}\omega_{i}\; \frac{\gamma_{\min}}{\gamma_i} \text{sgn}(\vec{v}_i \cdot\vec{v}_{\lambda})\vec{b}\cdot\vec{v}_{\lambda} \\ \nonumber
&=-\frac{2 \gamma_{\min}}{D } \sum_{i}  \frac{\omega_{i}}{\gamma_i} \sum_{\lambda \; s.t. \; \vec{v}_{\lambda} \cdot \vec{v}_{i} \geq0} \vec{v}_{\lambda}\cdot\vec{b} \nonumber \\ &=-\frac{2 \ell}{D}\gamma_{\min} \; \vec{a}\cdot\vec{b} \nonumber
\end{align}
where we have used equation \eqref{mod5} in the last step; see Appendix A for details. As for protocol 1, using equation \eqref{oppvec} we get that the marginals $\ave{a}=\ave{b}=0$. Hence the model reproduces the statistics of $\rho_W(\alpha)$ for $\alpha = \frac{2 \ell}{D} \gamma_{\text{min}}$. 

Starting from a sufficiently regular polyhedron with a large number $D$ of vertices $\vec{v}_{\lambda}$, we can approximate the unit sphere and the factor $\ell$ can become arbitrary close to one. In the limit $D\rightarrow\infty$ we expect to recover the uniform distribution over the sphere and our model therefore becomes equivalent to Werner's model for $\rho_{W}(1/2)$ \cite{werner89}. In Fig. I we plot upper bounds on the required shared randomness to simulate $\rho(\alpha)$ as a function of $\alpha$ obtained via protocol 2. We use a family of polyhedra, generated iteratively and starting from the icosahedron. To generate the second polyhedron, we take the union of the icosahedron and its normalized dual, and so on.

Note that the above protocols are LHS models. Hence the above results can be straightforwardly extended to the simulation of entangled states which are obtained via local filtering on the Werner state, \eg Ref. \cite{almeida07} (see Appendix A). Also, it would be interesting to see if more economical models (\ie using less shared randomness) exist, and if local entangled states require more shared randomness compared to separable states. For Werner states, this translates to whether we expect to see a discontinuity at the separable-entangled boundary for $\alpha=1/3$ (see Fig.I). We give two partial answers in this direction: (i) for LHS models, the maximum $\alpha$ one can simulate with $D=4$ is the separable state $\alpha=1/3$ (see Appendix B); (ii) Restricting to equatorial measurements one can achieve $\alpha=1/2$ with only $D=4$ (see Appendix C). 

\section{General results}

In the above, we have focused on a class of highly symmetric states, namely Werner states, and considered only projective measurements. Here we show how local models with finite shared randomness can be constructed for essentially any state that admits a LHV model. We also discuss the case of general measurements, \ie POVMs. 

{\bf Result 1.} Consider a state $\rho$ (of dimension $d \times d$) admitting a LHV model for all projective measurements. Then, a LHV model using only finite shared randomness can simulate all projective measurements on the state 
\small \ba \rho(\eta)= \eta ^2 \rho + \eta(1-\eta) \left(\frac{\mathbb{I}}{d}\otimes\rho_B  + \rho_A \otimes \frac{\mathbb{I}}{d} \right)+(1-\eta)^2 \frac{\mathbb{I} \otimes \mathbb{I}}{d^2}  \nonumber
\ea \normalsize
for any $0 \leq \eta<1$. Here $\rho_{A,B} = \Tr_{B,A}(\rho)$.

\begin{proof} First, note that it follows from the relation
\ba \label{res1} \tr[A_a\otimes B_b \rho(\eta) ] = \tr[A_a(\eta)\otimes B_b(\eta) \rho]  \ea
that the simulation of projective measurements (given by operators $A_a$ and $B_b$) on $\rho(\eta)$ is equivalent to the simulation of noisy measurements, given by operators $A_a(\eta)=\eta A_a + (1-\eta) \frac{\mathbb{I}}{d}$ and $B_b(\eta)=\eta B_b + (1-\eta) \frac{\mathbb{I}}{d}$ on the state $\rho$. Next, since $A_a(\eta)$ and $B_b(\eta)$ are full-rank for any $ \eta <1$, they can always be decomposed as convex mixtures over a single set of finitely many projective measurements \footnote{Note that Result 1 does not straightforwardly extend to the case of POVMs. First, notice that equation \eqref{res1} holds only for projective measurements. For general POVMs, one should define the noisy measurements as $A'_a(\eta) = \eta A_a + (1-\eta) \tr(A_a) \frac{\mathbb{I}}{d} $ and similarly for Bob. We believe that all $A'_a(\eta)$ can be decomposed over a single set of finitely many POVMs, although we have no formal proof.}. Finally, note that the simulation of a finite number of projective measurements on $\rho$ requires only finite shared randomness. This follows from the fact (i) the resulting distribution is local (as $\rho$ admits a LHV model), and (ii) the set of local distributions forms a polytope \cite{brunner_review}. 
\end{proof}

Note that the amount of shared randomness needed will depend on the value of $\eta$ and diverges as $\eta \rightarrow 1$.

{\bf Result 2.} Let us now consider more general measurements, \ie POVMs. Starting from an entangled state $\rho$ admitting a local model for projective measurements, the method developed in Ref. \cite{hirsch13} allows us to construct another entangled state $\rho'$ admitting a local model for POVMs. Importantly, the amount of shared randomness required in the final model coincides with that of the initial model. Hence starting from $\rho$ (of dimension $d \times d$) admitting a local model with $k$ bits of shared randomness for projective measurements, we get 

\ba \rho' = \frac{1}{(d+1)^2}  \left( \rho + d  (\rho_A \otimes F + F \otimes \rho_B)  + d^2  F \otimes F \right)  \nonumber \ea
which admits a local model for POVMs using also $k$ bits of shared randomness. Note that $F= \ket{d+1}\bra{d+1}$ denotes a projector onto a subspace orthogonal to the support of $\rho$, hence $\rho'$ is entangled by construction and of local dimension $d+1$.

Finally, we present two examples illustrating the above results. First, applying Result 1 the the local model of Ref. \cite{regev07} allows us to extend our result for two-qubit Werner states. Specifically we show that $\rho_{W}(\alpha)$ can be simulated with finite shared randomness for $\alpha \lesssim 0.66$. Upper bounds on the amount of shared randomness are given in Fig.1 (using again an iterative procedure based on the icosahedron). Notably, this shows that certain states useful for EPR steering can be simulated with finite shared randomness. Secondly applying Result 2 to the state\footnote{Note that we start here from a LHS model, hence the model works directly for POVMs on Bob's side. It is thus sufficient to apply the extension procedure of Result 2 on Alice's side only.} $\rho_{W}(0.43)$, we obtain that the state $\rho = \frac{1}{3} [\rho_W(0.43) + 2 \ket{2}\bra{2}\otimes \frac{\mathbb{I}}{2}]$ can be simulated for arbitrary POVMs using $\log_2(12)$ bits of shared randomness.

\section{Simulating nonlocal states with finite resources} 

Finally, we discuss the simulation of entangled states which are nonlocal. In this case, classical communication from (say) Alice to Bob is required. This communication is sent after Alice has received her input. Two cases can be considered: (i) Alice and Bob are initially uncorrelated (\ie have no shared randomness), and Alice sends classical information to Bob, (ii) Alice and Bob have access to shared randomness, and Alice sends classical information to Bob. Known protocols (see \eg \cite{toner03,regev07,branciard11}) require, for case (ii), finite communication assisted with infinite shared randomness---hence infinite communication for case (i). A notable exception is Ref. \cite{brassard13} which presents a model using no shared randomness and finite expected communication. Here we present protocols using only finite resources.

Considering case (i), we first show that the statistics of any bipartite entangled state that is full-rank can be simulated with finite communication. Specifically, consider a state of the form $\rho= \alpha \ket{\Psi}\bra{\Psi}+ (1-\alpha) \mathbb{I}/{d^2}$, where $\ket{\Psi}$ is an arbitrary entangled state of dimension $d \times d$, and $\alpha<1$. Upon receiving her measurement setting $A=\{A_{a}\}$, Alice outputs $a$ according to the distribution $p(a)=\Tr(\rho_{A}A_a)$ where $\rho_A$ is Alice's reduced state. For output $a$, Bob should hold the (normalized) state $\rho_B^a = \Tr_{A}(A_a \otimes \mathbb{I} \rho)/ p(a)$. Since $\rho_B^a$ is full-rank (by construction), then for any $\alpha<1$, there exists a polyhedron $V$ (with $D$ vertices, each representing a pure quantum state of dimension $d$) such that Alice can decompose $\rho_B^a$ as a convex combination of the vertices of $V$. With probability $\omega_i$ (the coefficient of vertex $i$ in the decomposition) Alice sends label $i$ to Bob, who can then locally reconstruct the corresponding pure state (knowing $V$). The model thus reproduces the statistics of $\rho$ using $\log_2(D)$ bits of communication.

For case (ii), we show that any state $\rho_W(\alpha)$ (see eq. \eqref{werner}), with $\alpha<1$, can be simulated with finite shared randomness and finite communication (worst case). In particular, for $\alpha \leq 3/4$ a single bit suffices. To construct such a model, we combine the ideas of Protocol 1 and the simulation model (using 1 bit of communication) for the singlet state of Ref. \cite{toner03}. See Appendix D for details.

\section{Conclusion}

We have shown that the correlations of essentially all entangled states that admit a LHV model can be simulated with finite shared randomness. This shows that the requirement of infinite shared randomness (hence channels with infinite capacity) used in previous models can in fact be dispensed with. 

An interesting open question is to find the minimal amount of shared randomness required to simulate a local entangled state? For a state of local dimension $d$, are more than $2\log_2(d)$ bits of shared randomness always required, that is, is the simulation of local entangled states strictly more costly than that of separable states. We presented a model using only 2 bits for Werner states of two-qubits, but our model works only for equatorial measurements, hence the question remains open. 

\emph{Acknowledgements.} We thank Antonio Acín and Marcin Pawłowski for discussions. We acknowledge financial support from the Swiss National Science Foundation (grant PP00P2\_138917 and QSIT), and SEFRI (COST action MP1006).




\providecommand{\href}[2]{#2}\begingroup\raggedright\endgroup



\begin{appendix}

\section{LHS models for Werner states using finite shared randomness} \label{werner_proj}

Here, we describe in detail the protocols 1 and 2 of the main text for the simulation of Werner states $\rho_W(\alpha)$ with $\alpha<1/2$. 

\emph{Protocol 1.} Consider $V$ to be any of the platonic solids (except for the tetrahedron) with $D$ vertices $\vec{v}_{i}$, satisfying conditions \eqref{oppvec} and \eqref{sumhs}. The shared variable is given by $\lambda \in \{1,\cdots,D\}$, uniformly distributed. Upon receiving $\lambda$, Alice and Bob use vector $\vec{v}_{\lambda}$, and output according to the following response functions:
\begin{align}
p_A(a|\lambda,\vec{a}) &=\frac{1 + a \text{sgn}(\vec{v}_{\lambda}\cdot\vec{a})}{2}, \label{arep} \\
p_B(b|\lambda,\vec{b}) &= \frac{1 - b  ( \vec{v}_{\lambda} \cdot \vec{b} )}{2}. \label{brep}
\end{align}
To begin with, consider the case where Alice's measurements corresponds to one of the vertices of $V$, \ie $\vec{a}=\vec{v}_{i}$. Bob's measurement $\vec{b}$ is arbitrary. We obtain the correlator: 
\begin{align}
\langle ab \rangle&=-\frac{1}{D}\sum_{\lambda} \text{sgn}(\vec{v}_i \cdot\vec{v}_{\lambda})\vec{y}\cdot\vec{v}_{\lambda} \nonumber\\
&=-\frac{1}{D} ( \sum_{\lambda|\vec{v}_{i}\cdot\vec{v}_{\lambda}\geq0}\vec{v}_{\lambda}\cdot\vec{b}-\sum_{\lambda|\vec{v}_{i}\cdot\vec{v}_{\lambda}<0}\vec{v}_{\lambda}\cdot\vec{b} \; ). \nonumber \\
\end{align}
From equation \eqref{oppvec} we have that 
\ba \sum_{\lambda|\vec{v}_{i}\cdot\vec{v}_{\lambda}\geq0}\vec{v}_{\lambda}=-\sum_{\lambda|\vec{v}_{i}\cdot\vec{v}_{\lambda}<0}\vec{v}_{\lambda} \ea 
hence implying that 
\begin{align}
\langle ab \rangle&=-\frac{2}{D}\sum_{\lambda|\vec{v}_{i}\cdot\vec{v}_{\lambda}\geq0}\vec{v}_{\lambda}\cdot\vec{b} =-\frac{2}{D} \gamma \; \vec{v}_{i}\cdot\vec{b} \nonumber 
\end{align}
where we used equation \eqref{sumhs} in the last step. Next we compute the marginals:
\begin{align}
\langle a \rangle &=-\frac{1}{D}\sum_{\lambda} \text{sgn}(\vec{v}_i\cdot\vec{v}_{\lambda}) = 0
\end{align}%
since for each $\vec{v}_j$ there is an opposite vector $\vec{v}_k = -\vec{v}_j$. Similarly for Bob:
\begin{align}
\langle b \rangle &=-\frac{1}{D}\sum_{\lambda} (\vec{b} \cdot \vec{v}_{\lambda}) = 0.
\end{align}
Hence the model simulates a Werner state for $\alpha =\frac{2 \gamma}{D}$, for the case in which Alice's measurement is one of the vertices of $V$.

Next we extend the model to an arbitrary projective measurement for Alice, represented by vector $\vec{a}$. Note that for any $\vec{a}$ one can find a set $\{\omega_{i}\}_{i=1,...,D}$, with $\omega_i\geq0$ and $\sum_{i=1}^{D}\omega_i=1$ such that 
\begin{align}
\sum_{i=1}^D\omega_i\vec{v}_{i}=\ell\vec{a},
\end{align}
with $\ell<1$. That is, for each $\vec{a}$ one can find a vector lying in the convex hull of $V$ that lies parallel to $\vec{a}$ and has length $\ell$. Hence, $\ell$ is precisely the radius of the sphere (centered in the origin) inscribed inside $V$.  If upon receiving $\vec{a}$ Alice uses local randomness to simulate the measurement of $\vec{v}_{i}$ with probability $\omega_{i}$ the overall correlator is given by
\begin{align}
\langle ab \rangle=\sum_{i=1}^D\omega_{i}\vec{v}_{i}\cdot\vec{b}=\ell\vec{a}\cdot\vec{b}.
\end{align}
The marginal remains unchanged, \ie $\langle a \rangle=0$. Hence, the model now simulates a Werner state $\rho_W(\alpha)$ with visibility $\alpha = \frac{2 \gamma \ell}{D}$. Indeed the 'shrinking factor' $\ell$ depends on the choice of polyhedron $V$.  


For each platonic solid, we give the visibility $\alpha$ of the Werner state that is simulated, and the required amount of shared randomness (see Table I). For the dodecahedron and the icosahedron, the model simulates the correlations of an entangled state. Note that the visibility $\alpha$ depends on the ratio of various parameters, hence using a polyhedron with more vertices may result in a lower visibility.

\begin{table}[t!] \label{table1}
 \begin{tabular}{|c||c|c|c|}
 \hline
  & Shared randomness (bits) & $\alpha$ & Sep/Ent \\
 \hline
 Octahedron & 2.58 &  0.19 & separable\\
 \hline
 Cube & 3 &  0.29 & separable\\
 \hline 
 Dodecahedron & 4.32 &  0.41 & entangled\\
 \hline
 Icosahedron & 3.58  & 0.43 & entangled\\
 \hline
 \end{tabular} \caption{For each platonic solid, we give the visibility $\alpha$ of the simulated Werner state. The amount of required shared randomness is given by the number of vertices of the polyhedron.}  
 \end{table} 

%
%

\emph{Protocol 2.} Protocol 1 can be extended to any polyhedron $V$ (with $D$ vertices) satisfying $\eqref{oppvec}$. Hence we now relax condition \eqref{sumhs}, and have the relation 
\ba   
\sum_{j \; s.t. \;  \vec{v}_{j}\cdot\vec{v}_{\lambda}\geq0}\vec{v}_{j}=\gamma_{\lambda} \vec{m}_{\lambda}
\ea
where $\vec{m}_{\lambda}$ is a normalized vector and generally $\vec{m}_{\lambda}\neq\vec{v}_{\lambda}$. Define $\gamma_{\text{min}} = \text{min}_{\lambda} ( \gamma_{\lambda} )$. Hence we obtain a second polyhedron $M$, defined by the vertices $\vec{m}_{\lambda}$, which are in one-to-one correspondence with the $\vec{v}_{\lambda}$. 

Upon receiving $\lambda \in \{1,...,D\}$, Alice and Bob use vector $\vec{v}_{\lambda}$. Similarly to above, let us start with the case where Alice's measurement corresponds to one of the vectors of $M$, $\vec{a}=\vec{m}_{i}$. Here we will slightly modify Alice's response function compared to protocol 1. Specifically, Alice now outputs according to 
\begin{align}
p_A(a|\lambda,\vec{m}_i) &=\frac{1 + a \text{sgn}(\vec{v}_{\lambda}\cdot\vec{v}_i)}{2},
\end{align}
with probability $\gamma_{\text{min}}/\gamma_{i}$, and outputs randomly otherwise. Bob receives an arbitrary projective measurement $\vec{b}$ and outputs as in protocol 1. The correlator is thus given by 
\begin{align}
\langle ab \rangle&=-\frac{1}{D}\sum_{\lambda} \frac{\gamma_{\text{min}}}{\gamma_i} \text{sgn}(\vec{v}_i \cdot\vec{v}_{\lambda})\vec{b}\cdot\vec{v}_{\lambda} \label{genmodel}\\
&=-\frac{1}{D} \frac{\gamma_{\text{min}}}{\gamma_i}  ( \sum_{\lambda|\vec{v}_{i}\cdot\vec{v}_{\lambda}\geq0}  \vec{v}_{\lambda}\cdot\vec{b}-\sum_{\lambda|\vec{v}_{i}\cdot\vec{v}_{\lambda}<0}\vec{v}_{\lambda}\cdot\vec{b} ) \nonumber \\
&=-\frac{2 \gamma_{\text{min}}}{D \gamma_i}\sum_{\lambda|\vec{v}_{i}\cdot\vec{v}_{\lambda}\geq0} \vec{v}_{\lambda}\cdot\vec{b} \nonumber \\
&=-\frac{2}{D}\gamma_{\text{min}} \; \vec{m}_{i}\cdot\vec{b}. \nonumber 
\end{align}
Note that condition \eqref{oppvec} again ensures that the marginals are uniform. Hence, the model simulates the correlations of a Werner state $\rho_W(\alpha)$ with visibility $\alpha=\frac{2}{D}\gamma_{\text{min}}$ when Alice's measurement corresponds to one of the vertices of $M$. Following the same reasoning as above (for protocol 1), we can extend the simulation model to the case of an arbitrary projective measurement for Alice. Similarly to above, the resulting visibility is found to be $\alpha = 2 \gamma_{\text{min}} \ell / D$, where $\ell$ is the radius of the sphere inscribed inside $M$ centered on the origin.

\emph{Extensions.} Protocols 1 and 2 are LHS models. Hence, they can be extended to the simulation of other entangled states, which can be obtained from Werner state via a filtering operation on Bob's side (the trusted party). For instance, Ref. \cite{almeida07} discussed states of the form 
\ba \rho_{\alpha, \theta} = \alpha \ketbra{\psi_{\theta}}{\psi_{\theta}} + (1- \alpha) \frac{\mathbb{I}}{2} \otimes \rho_B \ea
where $\ket{\psi_{\theta}} = \cos(\theta) \ket{00} + \sin(\theta) \ket{11}$ and $\rho_B = \Tr_A (\ketbra{\psi_{\theta}}{{\psi_{\theta}}} )$. Our model can be straightforwardly adapted to the above class of states. For a given amount of shared randomness, the model will simulate $\rho_{\alpha, \theta} $ with the same visibility $\alpha$ as for the Werner state.

\section{LHS models for entangled Werner states require more than two bits of shared randomness}

Consider LHS models. Bob's response function is quantum mechanical, given by the trace rule. 
In order to simulate a Werner state $\rho_W(\alpha)$ with such a model using only two bits of shared randomness, we must have that 
\begin{align} 
\langle ab \rangle &= \sum_{\lambda=1}^4 p_\lambda A_\lambda(\vec{a}) \; ( \vec{v}_{\lambda} \cdot \vec{b} ) = -\alpha \; \vec{a} \cdot \vec{b} .
\end{align}
where $\sum_\lambda p_\lambda = 1$ and $p_\lambda \geq 0$. Here $A_\lambda(\vec{a})$ denotes an arbitrary response function for Alice. For the particular case of $\vec{b} = \vec{a}$, this implies 
\begin{align} 
 \sum_{k=1}^4 p_k A_\lambda(\vec{a}) \; ( \vec{v}_{k} \cdot \vec{a} ) = -\alpha
\end{align}
hence we obtain
\begin{align} 
 \alpha \leq \max_{\lambda} ( | \vec{v}_{\lambda} \cdot \vec{a} |) .
\end{align}
As this holds for all $\vec{a}$, we have that 
\begin{align} 
\alpha \leq \min_{\vec{a}}  [ \max_{\lambda} ( | \vec{v}_{\lambda} \cdot \vec{a} |) ] .
\end{align}
Here the best strategy consists in using the tetrahedron, leading to $\alpha \leq 1/3$.
Consequently, any LHS model reproducing the correlations of an entangled Werner state, i.e. $\rho_W(\alpha)$ with $\alpha>1/3$, requires more than two bits of shared randomness.

\section{Simulating equatorial measurements on Werner states with two bits of shared randomness} \label{twobit_model}

Here we present a model to simulate the statistics of the state $\rho_{W}(\alpha)$ for $\alpha\leq1/2$, where all measurement Bloch vectors lie in a plane (taken here to be the $x-y$ plane). Surprisingly, the model only uses two bits of shared randomness. We parametrize Alice's and Bob's measurement vectors on the Bloch equator as $\vec{a}=(\cos(\theta_a),\sin(\theta_a))$ and $\vec{b}=(\cos(\theta_b),\sin(\theta_b))$. Imagine the following model which uses a single bit of shared randomness $\lambda=0,1$ with equal probability. For $\lambda=0$, Alice outputs according to the probability distribution 
\begin{align}
p_{A}(a|\lambda=0,\vec{a})=\frac{1}{2}(1+a\cos(\theta_a))
\end{align}
whereas for $\lambda=1$ she outputs according to 
\begin{align}
p_{A}(a|\lambda=1,\vec{a})=\frac{1}{2}(1+a\sin(\theta_a)).
\end{align}
Bob does exactly the same up to a flip of his output:
\begin{align}
p_{B}(b|\lambda=0,\vec{b})&=\frac{1}{2}(1-b\cos(\theta_b)) \\ 
p_{B}(b|\lambda=1,\vec{b})&=\frac{1}{2}(1-b\sin(\theta_b)).
\end{align}
A short calculation shows that this gives the correlator
\begin{align}
\langle ab \rangle&=-\frac{1}{2}(\cos(\theta_a)\cos(\theta_b)+\sin(\theta_a)\sin(\theta_b))\\
&=-\frac{1}{2}\vec{a}\cdot\vec{b}. \nonumber
\end{align}
In order to ensure that we have the correct marginals, we add an additional bit of shared randomness to the model $\mu=0,1$ (again uniform). If we have $\mu=1$ then Alice and Bob should both flip their output, \ie 
\begin{align}
p_A(a|\lambda=0,\mu,\vec{a})&=\frac{1}{2}(1+a(-1)^\mu\cos(\theta_a)); \\
p_B(a|\lambda=1,\mu,\vec{a})&=\frac{1}{2}(1+a(-1)^\mu\sin(\theta_a))
\end{align}
and equivalently for Bob. This then gives uniform marginals $\ave{a}=\ave{b}=0$ while keeping the correlator unchanged. Hence, we simulate exactly the statistics of projective equatorial measurements on the state $\rho_{W}(1/2)$. 

It would be interesting to see whether this model can be extended to the whole sphere. Using the techniques of Ref. \cite{brassard99}, we did not manage to solve the problem, as the visibility $\alpha$ is reduced in the procedure.

\section{Simulating nonlocal Werner states with finite communication and finite shared randomness} \label{comm_model}

We now discuss the simulation of a two-qubit Werner state $\rho_{W}(\alpha)$ for all $\alpha<1$ with finite communication and finite shared randomness. Consider a polyhedron $V$ with $D$ vertices satisfying \eqref{oppvec}, with corresponding $\gamma_{\text{min}}$ and shrinking factor $\ell$. Our model uses $n\log_2(D)$ bits of shared randomness and $\log_2{n}$ bits of communication (in the worst case), and simulates $\rho_{W}(\alpha)$ for
\begin{align}
\alpha=\frac{\gamma_{\min}}{\gamma_{\max}} \left( 1-[1- \frac{2\gamma_{\min}}{D}]^n \right) \ell^2
\end{align}
where $\gamma_{\max}= \max_i(\gamma_i)$. 
Note that by choosing a symmetric enough polyhedron with $\ell\approx1$ and $2\gamma_{\min}/D\approx 1/2$ we can simulate a $\rho_{W}(\alpha)$ for $\alpha \rightarrow 3/4$ with $n=2$ (when $D \rightarrow \infty$). Hence, using finite shared randomness and a single bit of communication suffices to simulate a nonlocal quantum state. 

Again we consider a polyhedron $V$ with $D$ vertices satisfying \eqref{oppvec} from which we can define a second polyhedron $M$ via equation \eqref{mod5}. Let us first discuss the case in which both Alice and Bob's measurement Bloch vectors correspond to one of the vertices of $M$, \ie Alice gets vector $\vec{a}=\vec{m}_{l}$ and Bob $\vec{b}=\vec{m}_{k}$. The protocol is then as follows:
\newline

{\bf Protocol 4.} In each round Alice and Bob receive $n$ numbers $\{\lambda_{1},\lambda_{2},\cdots,\lambda_{n}\}$, where each $\lambda_{i}$ is uniformly distributed with $\lambda_{i} \in \{1,\cdots,D\}$.  Either Alice will select one of the $\lambda_{i}$ or she will reject all of them. Consider a variable $T$ to denote Alice's selection, with $T=0$ corresponding to rejection. In the first step, Alice concentrates on $\lambda_{1}$ and does one of the following: (i) with probability $|\vec{m}_{l}\cdot\vec{v}_{\lambda_{1}}|\gamma_{\min}/\gamma_{\max}$ she selects $\lambda_1$ and sets $T=1$ and moves to the final step, (ii) with probability $1-|\vec{m}_{l}\cdot\vec{v}_{\lambda_{1}}|\gamma_{\min}/\gamma_{l}$ she discards $\lambda_1$ and moves to the second step (concentrating now on $\lambda_{2}$), (iii) she rejects, sets $T=0$ and moves to the final step. Hence, at step $j$ (if it is reached), Alice concentrates on $\lambda_j$. In the final step, Alice may have selected $\lambda_T$ or she may have rejected. In the case of rejection ($T=0$), Alice sends $c=1$ to Bob and outputs randomly. Otherwise, she sends $c=T$ to Bob and outputs $a=\text{sgn}[\vec{m}_{l}\cdot\vec{v_{\lambda_{T}}}]$. Bob then outputs $b=-\text{sgn}[\vec{v}_{k}\cdot\vec{v}_{\lambda_c}]$ with probability $\gamma_{min}/\gamma_{k}$, and otherwise outputs randomly. 
\newline

For the correlator we have
\small
\begin{align}
&\langle ab \rangle = \\
&-\frac{\gamma_{\min}}{D^{n}\gamma_{k}}\sum_{\{\lambda_{i}\}}\sum_{t=1}^{n}p(T=t|\{\lambda_{i}\},\vec{m}_{l})\text{sgn}(\vec{m}_{l}\cdot\vec{v}_{\lambda_{t}})\text{sgn}(\vec{v}_{k}\cdot\vec{v}_{\lambda_{t}})\nonumber\\
=&-\frac{\gamma_{\min}}{D^{n}\gamma_{k}}\sum_{t=1}^{n}\sum_{\lambda_t}^{D}\text{sgn}(\vec{m}_{l}\cdot\vec{v}_{\lambda_{t}})\text{sgn}(\vec{v}_{k}\cdot\vec{v}_{\lambda_{t}}) \nonumber
\\ &\;\;\;\;\;\;\;\;\;\;\;\;\;\;\;\;\;\;\;\;\;\;\;\; \times\sum_{\{\lambda_{i\neq t}\}}p(T=t|\{\lambda_i\},\vec{m}_{l}). \nonumber
\end{align}
\normalsize
From the protocol we have that
\small
\begin{align}
p(T=t|\{\lambda_i\},\vec{m}_j)=\frac{\gamma_{\min}}{\gamma_{\max}}|\vec{m}_{j}\cdot\vec{v}_{\lambda_t}|\prod_{j<t}(1-\frac{\gamma_{\min}}{\gamma_{j}}|\vec{m}_{j}\cdot\vec{v}_{\lambda_j}|).
\end{align}
\normalsize
From \eqref{genmodel} it follows that
\small
\begin{align}
&\frac{\gamma_{\min}}{\gamma_{k}}\sum_{\lambda_t}|\vec{m}_{l}\cdot\vec{v}_{\lambda_{t}}|\text{sgn}(\vec{m}_{l}\cdot\vec{v}_{\lambda_{t}})\text{sgn}(\vec{v}_{k}\cdot\vec{v}_{\lambda_{t}})\\
=&\frac{\gamma_{\min}}{\gamma_{k}}\sum_{\lambda_t}\vec{m}_{l}\cdot\vec{v}_{\lambda_{t}}\text{sgn}(\vec{v}_{k}\cdot\vec{v}_{\lambda_{t}}) \nonumber\\
=&2\gamma_{\min}\vec{m}_{l}\cdot\vec{m}_{k}. \nonumber
\end{align}
\normalsize
We then have
\small
\begin{align}
\langle ab \rangle &= \frac{-2\gamma^2_{\min}}{D^{n}\gamma_{\max}}\vec{m}_{l}\cdot\vec{m}_{k}\sum_{t=1}^{n}\sum_{\{\lambda_{i\neq t}\}}\prod_{j<t}(1-\frac{\gamma_{\min}}{\gamma_{l}}|\vec{m}_{l}\cdot\vec{v}_{\lambda_{j}}|)
\end{align}
and so we simulate a Werner state with $\alpha$ given by
\begin{align}
\alpha =  \frac{2\gamma^2_{\min}}{D^{n}\gamma_{\max}}\sum_{t=1}^{n}\sum_{\{\lambda_{i\neq t}\}}\prod_{j<t}(1-\frac{\gamma_{\min}}{\gamma_{l}}|\vec{m}_{l}\cdot\vec{v}_{\lambda_{j}}|).
\end{align}
We now proceed to simplify the above expression for $\alpha$ and show that it is independent of $\vec{m}_{l}$. Since each term in the product depends only on a single $\lambda_{j}$ we have:
\begin{align}
\alpha=\frac{2\gamma^2_{\min}}{\gamma_{\max}}\sum_{t=1}^{n}\frac{1}{D^{n-t+1}}\sum_{\{\lambda_{i>t}\}}\prod_{j<t}\sum_{\lambda=1}^{D}\frac{1}{D}(1-\frac{\gamma_{\min}}{\gamma_{l}}|\vec{m}_{l}\cdot\vec{v}_{\lambda_j}|).\nonumber
\end{align}
\normalsize
From the definition of $\gamma_{l}$, it follows that $\sum_{\lambda=1}^{D}\frac{1}{D}\frac{\gamma_{\min}}{\gamma_{l}}|\vec{m}_{l}\cdot\vec{v}_{\lambda_j}|=2\gamma_{\min}/D$, and so summing over the $\{\lambda_{i>t}\}$ as well we get
\begin{align}
\alpha&= \frac{2\gamma^2_{\min}}{D\gamma_{\max}}\sum_{t=1}^{n}\prod_{j<t}(1-2\gamma_{\min}/D)\\
&=\frac{2\gamma^2_{\min}}{D\gamma_{\max}}\sum_{t=1}^{n}(1-2\gamma_{\min}/D)^{t-1} \nonumber\\
&=\frac{\gamma_{\min}}{\gamma_{\max}}(1-(1-2\gamma_{\min}/D)^n), \nonumber
\end{align}
where in the last line we have used the fact that $\sum_{i=1}^{n}(1-x)^{(i-1)}=(1-(1-x)^n)/x$.
Using similar reasoning it is lengthy but straightforward to check that both Alice and Bob's marginals are uniform, \ie $\langle a \rangle = \langle b \rangle =0$. Finally, we note that we can extend this model to a model for all projective measurements in the same way as previously if Alice and Bob decompose their measurement vectors as convex combinations of vertices of the polyhedron $M$. This will add a factor $\ell^2$ giving the final visibility

\begin{align}
\alpha=\frac{\gamma_{\min}}{\gamma_{\max}} \left( 1-[1- \frac{2\gamma_{\min}}{D}]^n \right) \ell^2.
\end{align}

\begin{figure}[t!]
\includegraphics[width = \columnwidth]{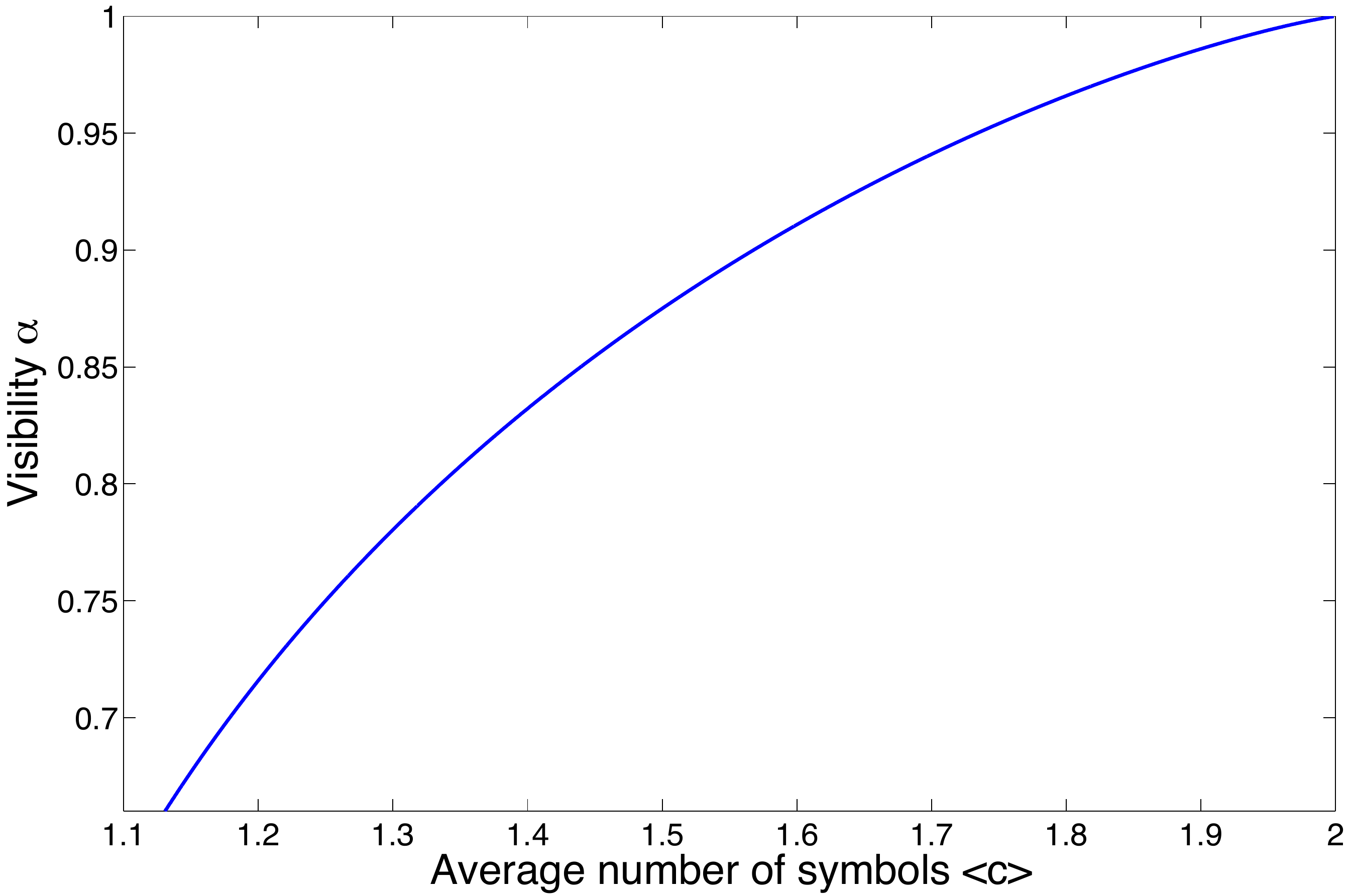}
    \caption{Visibility $\alpha$ of the Werner state $\rho_W(\alpha)$ as a function of the average number of communication $\ave{c}$.}
\label{Fig2}
\end{figure}

\emph{Average communication.} Although the above protocol requires $\log_{2}(n)$ bits of communication in the worst case, the average amount of communication is typically much smaller, as each $\lambda_{i}$ is decreasingly less likely to be selected by Alice. To quantify this, we calculate the average label that is sent by Alice. \ie the average value of the communication $c$:
\begin{align}
\ave{c} &= \frac{1}{D^n}\sum_{\{\lambda_i\}}\sum_{j=1}^{n} j \,p(c=j | \{ \lambda_{i} \} \vec{a}) \\
&= 1 + (1-x)x \frac{d}{dx} g_n(x)\nonumber
\end{align}
with $x=1-2\gamma_{\min}/D$ and $g_n(x)= \frac{1-x^n}{1-x}$. In the limit of large shared randomness, \ie $D \rightarrow \infty$, we get 
\begin{align}
\ave{c} &= 2- \frac{1+n}{2^n}.
\end{align}
Hence, in this regime, the average value of the communication $c$ remains smaller than 2. Figure 2 shows the visibility $\alpha$ of the simulated state as a function of $\ave{c}$. Thus we expect that the model requires only a small amount of average bits of communication although the worst case communication is $\log_2{n}$ bits.


\end{appendix}

\end{document}